\documentclass[twocolumn,showpacs,nofootinbib]{revtex4}
\usepackage{amssymb,eucal,amsthm,epsf,graphics,float}
\usepackage[english]{babel}
\def\e{{\mathop{\,\rm e}}}
\def\ii{\dot{\rm\i\hspace{-1.7pt}\i}}
\def\Z{\mathbb{Z}}

\def\vf{v_{\!f}}
\begin{document}
\title{Functional Renormalization Group calculation of the Fermi
surface of a spin-ladder}
\author{G. Abramovici}
\affiliation{Laboratoire de Physique des Solides,
Univ. Paris-Sud, CNRS, UMR 8502, F-91405 Orsay Cedex France}
\author{M. H\'eritier}
\affiliation{Laboratoire de Physique des Solides,
Univ. Paris-Sud, CNRS, UMR 8502, F-91405 Orsay Cedex France}
\begin{abstract}
We study non conventional superconductivity on a ladder, improving the
predictions of the Hubbard model. The determination of the Fermi surface,
in 2 or 3 dimensions, remains a very hard task, but it is exactly solvable for
a single ladder. We use functional Renormalisation Group methods, which prove,
here, scheme-dependant. In the superconducting phase, the
binding/antibinding gap is stabilized, but in the antiferromagnetic phase, it
shrinks and the ladder turns one-dimensional.
\end{abstract}
\maketitle
\section{Introduction}

The mechanism of unconventional superconductivity remains a major source of
interest for theorists. Anisotropic crystals, in particular organic ones,
are expected, both from a theoretical point of view and experimentally, to
show unconventional behaviour. Among the many materials, which are currently
studied, $Sr_{14-x} Ca_x Cu_{24} O_{41}$\cite{Azuma,Carron}
or $(La Y)_y(Sr Ce)_{14-y} Cu_{24} O_{41}$\cite{Hiroi,Sato} show very
anisotropic structures, and can be well represented by a single ladder. 
Indeed, unconventional superconductivity has been observed in these materials%
\cite{Dagotto,Piskunov}; in particular, superconducting (SC) phases are
found in the vicinity of antiferromagnetic ones, sometimes with a gapped
phase in between\cite{Lehur}. These observations indicate the possibility of
coexistence of magnetism and superconductivity.

The materials we have studied in this work belong to the class of strongly
correlated quasi-one-dimensional systems of electrons. One important
peculiarity of these systems is the existence of nearly degenerate phases with
different symmetries and, therefore, the possibility of very rich phase
diagrams, commonly observed experimentally and often predicted theoretically.
In quasi-one-dimensional systems, it is well established that two kinds of
Fermi surface instabilities can occur: (i) Cooper instabilities, including
singlet or triplet superconducting ones; (ii) Peierls instabilities, including
Spin Density Wave (SDW) or Charge Density Wave (CDW) ones. A very important
and quite general property is that these two kinds of instability are coupled,
because of the topology of the phase space in one dimension. The questions
about how these instabilities compete and sometimes coexist are very important
for the physics of these systems. Because of this instability coupling, such
a discussion cannot be done in a Fermi liquid approach, but requires more
sophisticated methods.

Different approaches have been used, to understand unconventional
superconductivity, like Monte Carlo simulations with the $t{-}J$ model%
\cite{Kimura, Riera} or with the Hubbard model\cite{Randeria,Bulut},
or exact diagonalization\cite{Germain}, DMRG method\cite{Noack}
or variational approach\cite{Sierra}, etc.\cite{Moreo}.

An important step forward has been taken with the use of the Renormalization
Group (RG)\cite{Wilson}. Not only can these calculations predict a SC phase%
\cite{Kuroki,Schulz}, but they give a new interpretation of this unconventional
mechanism: it results from the competition between the Cooper channel
(formation of pairs of electrons) and the Peierls channel (formation of
of electron-hole pairs)\cite{Giamarchi}. 

The RG method is a fixed point method, its application in condensed matter has a
severe drawback: the RG flow is always diverging, so that no
exact fixed point can be obtained; in other words, it is impossible to
calculate the renormalized parameters of these systems. Nevertheless, one can
calculate the phase diagram, by examining which susceptibilities are diverging
(i.e. are unstable) and which are remaining finite: the processes corresponding
to non divergent susceptibilities are negligible compared to those corresponding
to divergent ones.

This paper is devoted to the study of a ladder system, which consists of two
coupled chains of atoms, the intrachain coupling is written $t_\Vert$, the
interchain one $t_\perp$, with $t_\perp\ll t_\Vert$. We use the Hubbard model,
which has been widely studied by theorists\cite{Hubbard}, though its complete
analysis hasn't yet been achieved. M. Fabrizio\cite{Fabrizio} has previously
calculated the phase diagram of the ladder by a two-loop expansion using RG
equations. He obtains a very rich diagram, with Hubbard parameter $U$ ranging
over $[0,18\pi \vf]$ ($\vf$ is the Fermi velocity) and $t_\perp$ over
$[0,1.2\Lambda_o]$ ($\Lambda_o$ is the half band width), though its validity is
somehow questionable, since $U/\vf$ is the parameter of this expansion. If one
focuses on range $U\in[0,2\pi \vf]$, Fabrizio predicts a superconducting phase,
which he named phase~I; in this phase, the RG flow of susceptibilities shows
several divergences: the SDW channel coexists with the superconducting one. 

We proved recently, for small values of the interchain interaction $t_\perp$,
the existence of an extra SDW phase, in this region of parameters, by
including $K_\Vert$ dependence of the couplings\cite{Abramovici}. This phase is
characterized by the flow of all superconducting susceptibilities, which remain
finite, while SDW ones diverge. These calculations have been performed with a
fixed Fermi surface. This work also established the importance of high
energy processes (like the backward interband scattering $g_b$, see below)
during the RG flow: although these processes die before the flow becomes
divergent, they prove eventually influential.

In this paper, we will discuss the effect of the renormalization of the Fermi
surface, in the line of these $K_\Vert$ dependant RG calculations. One of
the questions is whether our results, in particular the existence of a SDW phase,
are valid or not. The answer is fortunately yes.

In the last decade, RG methods have achieved very sophisticated schemes: here,
we use either the One Particle Irreducible (OPI) scheme, following 
Ref.~\cite{Metzner,Honerkamp}, or the Wick-ordered one, following
Ref.~\cite{Salmhofer}, and calculated the scatterings in a one-loop expansion.
The renormalization of the Fermi surface remains valid, in this approximation.

A remarkable result is that the phase diagram becomes scheme-dependant. This
question was first addressed by H. Schulz, who argued that high energy processes
would be influent in specific cases: this implies that the way they are
included in the RG calculation would matter\cite{Schulz}. The response it
receives here contradicts the usual opinion, shared by a number of specialists,
that all schemes are equivalent and give identical results.

We will first describe the model (section \ref{modele}) and the RG equations
(section \ref{equationsRG}), then discuss the choice of the RG scheme
(section \ref{choixschema}) and analyse our results (section \ref{resultats}).

\section{Model}
\label{modele}

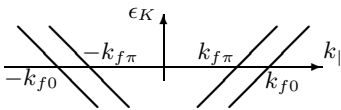
\begin{figure}[b]
\begin{center}
\begin{picture}(120,40)
\put(0,15){\vector(1,0){120}}
\put(60,10){\vector(0,1){25}}
\put(125,20){\makebox(0,0){$k_\Vert$}}
\put(10,10){\makebox(0,0){$-k_{f0}$}}
\put(40,20){\makebox(0,0){$-k_{f\pi}$}}
\put(80,20){\makebox(0,0){$k_{f\pi}$}}
\put(105,9){\makebox(0,0){$k_{f0}$}}
\put(52,35){\makebox(0,0){$\epsilon_K$}}
\thicklines
\put(5,30){\line(1,-1){30}}
\put(17,30){\line(1,-1){30}}
\put(103,30){\line(-1,-1){30}}
\put(115,30){\line(-1,-1){30}}
\end{picture}
\end{center}
\caption{The 2-band dispersion in $\Vert$ direction}
\label{dispersion}
\end{figure}

In a ladder, there are two separated bands in the dispersion diagram (0:
binding and $\pi$: antibinding), because of the Coulombian interaction between
the chains. In other words, in the $K_\perp$ direction, there are only two
physical points, $O$ and $\pi/b$ ($b$ is the interchain distance). There
are four Fermi points ($-k_{f0}$, $-k_{f\pi}$, $k_{f\pi}$, $k_{f0}$) in the
$K_\Vert$ direction (see Fig.~\ref{dispersion}). We will simply note $K$, for
the momenta in the $K_\Vert$ direction (and $k$ will always be the relative
momentum to a given Fermi point).

The Fermi surface gap is defined as $\Delta k_f\equiv k_{f0}-k_{f\pi}$. From
Luttinger theorem, $k_{f0}+k_{f\pi}$ is constant, so $\Delta k_f$ is the
only Fermi surface parameter. It relates $t_\perp$ the interchain interaction
by $\Delta k_f=2t_\perp/\vf$. 

The kinetic Hamiltonian is linearized around the Fermi points\cite{Solyom}
with a single Fermi velocity $\vf$, and writes ($R$ reads \textit{right moving}
particle and $L$ \textit{left moving} one):
\begin{eqnarray*}
&&H_{\rm cin}=
\sum_\sigma \vf\Big(\sum_K(K-k_{f0})R^\dag_{0\sigma}(K)R_{0\sigma}(K)+\\
&&(K-k_{f\pi})R^\dag_{\pi\sigma}(K)R_{\pi\sigma}(K)
+(K+k_{f0})L^\dag_{0\sigma}(K)L_{0\sigma}(K)\\&&\qquad+
(K+k_{f\pi})L^\dag_{\pi\sigma}(K)L_{\pi\sigma}(K)\Big).
\end{eqnarray*}

\begin{figure}[t]
\begin{center}
\newsavebox{\croix}
\begin{picture}(220,190)
\savebox{\croix}(20,20)[h]{%
\put(0,20){\line(1,-1){20}}
\put(0,0){\line(1,1){20}}
\put(7,13){\vector(1,-1){0}}
\put(17,3){\vector(1,-1){0}}
\put(7,7){\vector(1,1){0}}
\put(17,17){\vector(1,1){0}}}
\put(55,160){\usebox{\croix}}
\put(30,185){\makebox{$\scriptscriptstyle k_{f0}+k_1$}}
\put(25,155){\makebox{$\scriptscriptstyle -k_{f0}+k_2$}}
\put(75,185){\makebox{$\scriptscriptstyle -k_{f0}+k'_2$}}
\put(75,155){\makebox{$\scriptscriptstyle k_{f0}+k'_1$}}
\put(0,170){\makebox{$g_0\ :$}}

\put(55,110){\usebox{\croix}}
\put(30,135){\makebox{$\scriptscriptstyle k_{f0}+k_1$}}
\put(25,105){\makebox{$\scriptscriptstyle -k_{f\pi}+k_2$}}
\put(75,135){\makebox{$\scriptscriptstyle -k_{f\pi}+k'_2$}}
\put(75,105){\makebox{$\scriptscriptstyle k_{f0}+k'_1$}}
\put(0,120){\makebox{$g_{\!f0}\ :$}}

\put(55,60){\usebox{\croix}}
\put(30,85){\makebox{$\scriptscriptstyle k_{f0}+k_1$}}
\put(25,55){\makebox{$\scriptscriptstyle -k_{f0}+k_2$}}
\put(75,85){\makebox{$\scriptscriptstyle -k_{f\pi}+k'_2$}}
\put(75,55){\makebox{$\scriptscriptstyle k_{f\pi}+k'_1$}}
\put(0,70){\makebox{$g_{t0}\ :$}}

\put(55,10){\usebox{\croix}}
\put(30,35){\makebox{$\scriptscriptstyle k_{f0}+k_1$}}
\put(25,5){\makebox{$\scriptscriptstyle -k_{f\pi}+k_2$}}
\put(75,35){\makebox{$\scriptscriptstyle -k_{f0}+k'_2$}}
\put(75,5){\makebox{$\scriptscriptstyle k_{f\pi}+k'_1$}}
\put(0,20){\makebox{$g_{b0}\ :$}}

\put(175,160){\usebox{\croix}}
\put(150,185){\makebox{$\scriptscriptstyle k_{f\pi}+k_1$}}
\put(145,155){\makebox{$\scriptscriptstyle -k_{f\pi}+k_2$}}
\put(195,185){\makebox{$\scriptscriptstyle -k_{f\pi}+k'_2$}}
\put(195,155){\makebox{$\scriptscriptstyle k_{f\pi}+k'_1$}}
\put(120,170){\makebox{$g_\pi\ :$}}

\put(175,110){\usebox{\croix}}
\put(150,135){\makebox{$\scriptscriptstyle k_{f\pi}+k_1$}}
\put(145,105){\makebox{$\scriptscriptstyle -k_{f0}+k_2$}}
\put(195,135){\makebox{$\scriptscriptstyle -k_{f0}+k'_2$}}
\put(195,105){\makebox{$\scriptscriptstyle k_{f\pi}+k'_1$}}
\put(120,120){\makebox{$g_{\!f\pi}\ :$}}
\put(175,60){\usebox{\croix}}
\put(150,85){\makebox{$\scriptscriptstyle k_{f\pi}+k_1$}}
\put(145,55){\makebox{$\scriptscriptstyle -k_{f\pi}+k_2$}}
\put(195,85){\makebox{$\scriptscriptstyle -k_{f0}+k'_2$}}
\put(195,55){\makebox{$\scriptscriptstyle k_{f0}+k'_1$}}
\put(120,70){\makebox{$g_{t\pi}\ :$}}

\put(175,10){\usebox{\croix}}
\put(150,35){\makebox{$\scriptscriptstyle k_{f\pi}+k_1$}}
\put(145,5){\makebox{$\scriptscriptstyle -k_{f0}+k_2$}}
\put(195,35){\makebox{$\scriptscriptstyle -k_{f\pi}+k'_2$}}
\put(195,5){\makebox{$\scriptscriptstyle k_{f0}+k'_1$}}
\put(120,20){\makebox{$g_{b\pi}\ :$}}
\end{picture}
\end{center}
\caption{Schematic definitions of the couplings $\cal G$}
\label{couplings}
\end{figure}
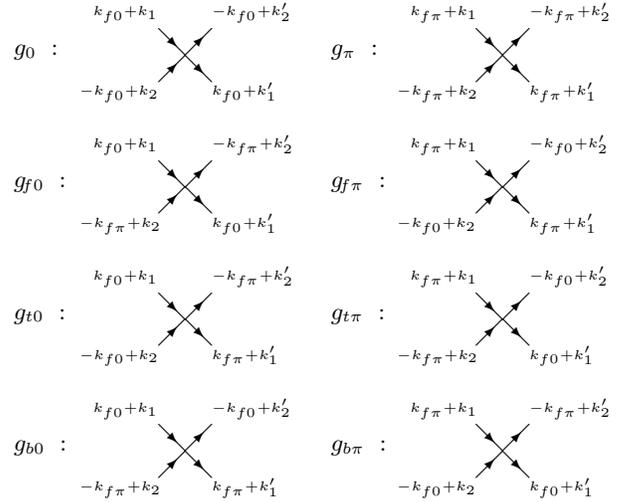

The interaction Hamiltonian writes 
\begin{eqnarray*}
&&H_{\rm int}=
{1\over N}\!\!\!\!\sum_{{\bf K}_1,{\bf K}_2,{\bf K}'_1,{\bf K}'_2
\atop{\bf K}_1+{\bf K}_2={\bf K}'_1+{\bf K}'_2}\sum_{\sigma_1,\sigma_2}
{\cal G}_4R_{{\bf K}'_1\sigma_1}^\dag R_{{\bf K}'_2\sigma_2}^\dag\times\\
&&R^{\phantom{\dag}}_{{\bf K}_2\sigma_2}R^{\phantom{\dag}}_{{\bf K}_1\sigma_1}+
{\cal G}_4L_{{\bf K}'_1\sigma_1}^\dag L_{{\bf K}'_2\sigma_2}^\dag
L^{\phantom{\dag}}_{{\bf K}_2\sigma_2}L^{\phantom{\dag}}_{{\bf K}_1\sigma_1}\\
&&+{\cal G}_1({\bf K}_1,{\bf K}_2,{\bf K}'_2,{\bf K}'_1)
R_{{\bf K}'_1\sigma_1}^\dag L_{{\bf K}'_2\sigma_2}^\dag
R^{\phantom{\dag}}_{{\bf K}_2\sigma_2}L^{\phantom{\dag}}_{{\bf K}_1\sigma_1}\\
&&+{\cal G}_2({\bf K}_1,{\bf K}_2,{\bf K}'_2,{\bf K}'_1)
R_{{\bf K}'_1\sigma_1}^\dag L_{{\bf K}'_2\sigma_2}^\dag
L^{\phantom{\dag}}_{{\bf K}_2\sigma_2}
R^{\phantom{\dag}}_{{\bf K}_1\sigma_1}
\end{eqnarray*}
in which ${\cal G}_\alpha$ is the two-particle coupling, and we have used
the g-ology representation. More precisely, there are 8 different couplings
$g_0$, $g_\pi$, $g_{f0}$, $g_{f\pi}$, $g_{t0}$, $g_{t\pi}$, $g_{b0}$ and
$g_{b\pi}$, corresponding to the interaction processes shown in
Fig.~\ref{couplings} (in this way, all $K_\perp$ dependence of the couplings
is included in the symbolic names, whereas all $K_\Vert$ dependence is
given in their arguments, see more details in Ref.~\cite{Abramovici}), plus the 
${\cal G}_4={\cal G}(RRRR)={\cal G}(LLLL)$ couplings which are not
renormalized in a one-loop expansion. At the beginning of the RG flow
($\Lambda=\Lambda_0$), all scatterings ${\cal G}_\alpha$ are set to $U$, the
Hubbard constant, thus one simply gets ${\cal G}_4=U$.

\section{RG equations}
\label{equationsRG}

The RG equations for scatterings $g_0$, $g_\pi$, $g_{f0}$, ..., $g_{b\pi}$
express their derivative as the sum of two terms: the first term is usually
called Cooper term, since it comes from an electron-electron diagram; the second
one is called Peierls term, since it comes from an electron-hole diagram. One
can write, in a generic way,
\begin{eqnarray}
&&\qquad\qquad
{\partial{\cal G}\over\partial\ell}({\bf K}_1,{\bf K}_2,{\bf K}_3,{\bf K}_4)
=\nonumber\\
&{\cal C}&\sum_{\bf K',K''}
{\cal G}({\bf K}'_1,{\bf K}'_2,{\bf K}'_3,{\bf K}'_4)
{\cal G}({\bf K}''_1,{\bf K}''_2,{\bf K}''_3,{\bf K}''_4)+\nonumber\\
&{\cal P}&\sum_{\bf K',K''}
{\cal G}({\bf K}'_1,{\bf K}'_2,{\bf K}'_3,{\bf K}'_4)
{\cal G}({\bf K}''_1,{\bf K}''_2,{\bf K}''_3,{\bf K}''_4)
\label{RGscattering}
\end{eqnarray}
in which $\cal C$ and $\cal P$ are coefficients ($\cal C$ stands for Cooper term
while $\cal P$ stands for Peierls term); explicit and detailed sums are given in
Appendix \textbf{B} of Ref.~\cite{Abramovici}; $\ell$ is the flow parameter (the
half band width is $\Lambda=\Lambda_o\e^{-\ell}$).

For all couplings, except $g_{b0}$ and $g_{b\pi}$, we get, in the OPI scheme,
${\cal C}=1/(4+2|\tilde K_1+\tilde K_2|)$ and, in the Wick-ordered one,
${\cal C}=1/(4-2|\tilde K_1+\tilde K_2|)$ (here $\tilde K\equiv\vf K/\Lambda$);
we get, in the OPI scheme,
${\cal P}=1/(4+2|\tilde K_1-\tilde K_3|)$ and, in the Wick-ordered one,
${\cal P}=1/(4-2|\tilde K_1-\tilde K_3|)$. In fact, the generic expression
(\ref{RGscattering}) does not apply to couplings $g_b$: the Cooper term
splits into two terms, one with the same ${\cal C}$ factor, one with a special
factor ${\cal C}^{\rm sp}=1/(4+2|\tilde K_1+\tilde K_2\pm2\Delta \tilde k_f|)$
for the OPI scheme and ${\cal C}^{\rm sp}=
1/(4-2|\tilde K_1+\tilde K_2\pm2\Delta \tilde k_f|)$ for the Wick-ordered one
($\Delta\tilde k_f\equiv \vf\Delta k_f/\Lambda$);
the Peierls term splits into two terms, one with the same
${\cal P}$ factor, one with a special factor
${\cal P}^{\rm sp}=1/(4+2|\tilde K_1-\tilde K_3\pm2\Delta \tilde k_f|)$ for the
OPI scheme and
${\cal P}^{\rm sp}=1/(4-2|\tilde K_1-\tilde K_3\pm2\Delta \tilde k_f|)$ for
the Wick-ordered one ($\pm$ reads $+$ for $g_{b0}$ and $-$ for $g_{b\pi}$).

The RG equation for $\Delta k_f$ is obtained through the two-loop expansion of
the self-energy $\Sigma$, following a standard calculation%
\cite{Fabrizio,Dupuis,Doucot}. Let
$G_{\rm o}=Z/(-\ii\omega+\vf(K-k_{f\theta}+\mu))$ be the free Right propagator
of the band $\theta$ ($\theta=0,\pi$), and $\mu$ the chemical potential, one can
write
$$
\Sigma_{R\theta}=\delta G_{\rm o}^{-1}={1\over Z}
(\delta \vf(K-k_{f\theta})-\vf\delta k_{f\theta}+\delta\mu)-
{G_{\rm o}^{-1}\over Z}\delta Z\ .
 $$

\begin{figure}
\begin{picture}(50,80)
\put(0,0){\line(1,0){50}}
\multiput(25,0)(0,5){8}{\line(0,1){2}}
\put(25,0){\circle*{2}}
\put(25,38){\circle*{2}}
\put(25,50){\circle{25}}
\put(12,0){\vector(1,0){0}}
\put(38,0){\vector(1,0){0}}
\put(28,62){\vector(1,0){0}}
\put(12,8){\makebox(0,0){$K$}}
\put(25,70){\makebox(0,0){$K'$}}
\put(38,8){\makebox(0,0){$K$}}
\end{picture}
\caption{One-loop tadpole diagram}
\label{oneloop}
\end{figure}
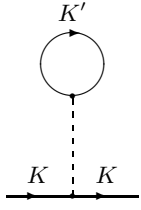

In Fig.~\ref{oneloop}, we show the tadpole diagram, corresponding to a one-loop
contribution in this expansion of $\Sigma$:
after all simplifications (one has to subtract carefully the contribution of
$\delta\mu$), one gets, in this one-loop expansion,
\begin{eqnarray*}
&&\delta k_{f0}=-\delta k_{f\pi}
={Z\over\pi \vf^2}\times\\
&&\left(g_{\pi2}-g_{02}+
g_{f\pi2}-g_{f02}-{g_{\pi1}-g_{01}+g_{f\pi1}-g_{f01}\over2}\right)
\end{eqnarray*}

It is obvious, in this formula, that $\Delta k_f$ depends on $K_\Vert$,
however, this dependency gives very small variations and can be
neglected.

\begin{figure}
\begin{picture}(50,80)
\put(0,0){\line(1,0){50}}
\multiput(25,0)(0,4){2}{\line(0,1){3}}
\multiput(25,31)(0,4){2}{\line(0,1){3}}
\put(25,0){\circle*{2}}
\put(25,7){\circle*{2}}
\put(25,31){\circle*{2}}
\put(25,38){\circle*{2}}
\put(25,50){\circle{25}}
\put(25,19){\circle{25}}
\put(12,0){\vector(1,0){0}}
\put(38,0){\vector(1,0){0}}
\put(13,20){\vector(0,1){0}}
\put(0,80){\makebox(0,0){$(a)$}}
\put(25,22){\makebox(0,0){$K'$}}
\put(28,62){\vector(1,0){0}}
\put(25,70){\makebox(0,0){$K''$}}
\end{picture}
\begin{picture}(80,100)
\put(0,30){\line(1,0){73}}
\put(19,30){\circle*{2}}
\put(51,30){\circle*{2}}
\put(35,30){\circle{30}}
\put(35,30){\vector(1,0){0}}
\put(38,46){\vector(1,0){0}}
\put(32,14){\vector(-1,0){0}}
\put(7,30){\vector(1,0){0}}
\put(20,80){\makebox(0,0){$(b)$}}
\put(5,38){\makebox(0,0){$K$}}
\put(68,30){\vector(1,0){0}}
\put(65,38){\makebox(0,0){$K$}}
\put(34,24){\makebox(0,0){$K'''$}}
\put(30,7){\makebox(0,0){$K''$}}
\put(30,55){\makebox(0,0){$K'$}}
\end{picture}
\caption{Two-loop tadpole diagram}
\label{twoloop}
\end{figure}
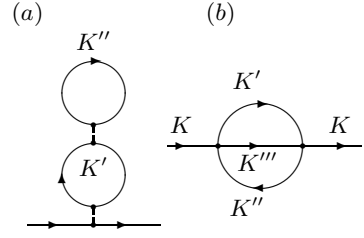

There are two different two-loop diagrams, represented in Fig.~\ref{twoloop}.
The first one $(a)$ gives no contribution, and the second one $(b)$ (sunrise)
gives three.

Two of them give logarithmic terms; in fact, these two contributions can be
deduced one from the other using symmetry $AA'$ (see notations in
Ref.~\cite{Abramovici}). 

The only contribution containing ${\cal G}_4$ is
\begin{eqnarray*}
&&\delta(\Delta k_f)\rfloor_{\rm 2-loop\atop{\cal G}_4}=
-{4ZU^2\delta\Lambda\over\pi^2\vf^3}\times\\
&&\left(
\int\limits_{-\Lambda+2\vf\Delta k_f}^0\!\!\!\!\!\!\!\!
{d\epsilon\over-2\Lambda+2\vf\Delta k_f}
-\int\limits_{-\Lambda-2\vf\Delta k_f}^0\!\!\!\!\!\!\!\!
{d\epsilon\over-2\Lambda-2\vf\Delta k_f}
\right.\\&&
+\left.
\int\limits_0^{\Lambda+2\vf\Delta k_f}\!\!\!\!\!\!\!\!
{d\epsilon\over2\Lambda+2\vf\Delta k_f}
-\int\limits_0^{\Lambda-2\vf\Delta k_f}\!\!\!\!\!\!\!\!
{d\epsilon\over2\Lambda-2\vf\Delta k_f}
\right)
\\&&
=-{4ZU^2\delta\Lambda\over\pi^2\vf^3}\times\left\{\matrix{
{1+2\vf\Delta k_f/\Lambda\over1+\vf\Delta k_f/\Lambda}
\hbox{ if }\Lambda\le2\vf\Delta k_f\cr
{2\vf\Delta k_f/\Lambda\over1-\vf^2\Delta k_f^2/\Lambda^2}
\hbox{ if }\Lambda\ge2\vf\Delta k_f\cr}\right.
\end{eqnarray*}

We have calculated the complete expression of the two-loop expansion, including
$K_\Vert$ dependence, both in the OPI scheme and in the Wick-ordered one. Except
for the last ${\cal G}_4$ contribution, these expressions  depend on the RG
scheme. It would be too fastidious to explicit all contributions: instead, let
us skip all $K_\Vert$ dependence. We get the following contribution:
\begin{eqnarray*}
&&\delta(\Delta k_f)\rfloor_{\rm 2-loop\atop except\ {\cal G}_4}
=\mp{4Z\delta\Lambda\over\pi^2\vf^3}\left(
{\vf k-\Lambda\over|\vf k-\Lambda|}+{\vf k+\Lambda\over|\vf k+\Lambda|}+
\right.\\
&&\left.2\log({-2\Lambda_{\rm o}-\vf k\over-\Lambda-|\Lambda-\vf k|})
+2\log({2\Lambda_{\rm o}-\vf k\over\Lambda+|\Lambda+\vf k|})\right)\times\\
&&\left(({\cal G}_1)^2+({\cal G}_2)^2-{\cal G}_1{\cal G}_2\right)
\end{eqnarray*}
for each $g_0$, $g_{f0}$, $g_{t0}$ (for which $\pm$ reads $+$),
$g_\pi$, $g_{f\pi}$, and $g_{t\pi}$ (for which $\pm$ reads $-$).
This is similar to previous calculations (Ref.~\cite{Rohe} for the OPI scheme,
Ref.~\cite{Katanin} for the Wick-ordered one), but we would like to emphasize
one major novelty: for $g_{b0}$ and $g_{b\pi}$, the second factor is
modified and writes
\begin{eqnarray*}
&&\!\!\!\!\!\!\!
{\vf(k\pm2\Delta k_f)-\Lambda\over|\vf(k\pm2\Delta k_f)-\Lambda|}+
{\vf(k\pm2\Delta k_f)+\Lambda\over|\vf(k\pm2\Delta k_f)+\Lambda|}+
\\
&&\!\!\!\!\!\!\!
2\log({-2\Lambda_{\rm o}-\vf(k\mp2\Delta k_f)
\over-\Lambda-|\Lambda-\vf(k\mp2\Delta k_f)|}\;
{2\Lambda_{\rm o}-\vf(k\mp2\Delta k_f)
\over\Lambda+|\Lambda+\vf(k\pm2\Delta k_f)|})
\end{eqnarray*}
in which the $\pm$ reads as in the first factor.

To end with technical details, let us explain the approximations used in the RG
equations. First, all scattering $\cal G$ depend on three arguments
$(k_1,k_2,k_3)$, which are replaced by their $2p_i\Delta
k_f$ ($p_i\in\Z$) approximation. This is generalized to all other couplings.
Second, the list of all functional couplings ${\cal G}((k_1,k_2,k_3)$ is
truncated by setting $|p_i|=2$, 3 or 4. Extra couplings are replaced by the
closer element in the list using symmetry preserving relations (cf.
Ref.~\cite{Abramovici}). Last, couplings ${\cal G}((k_1,k_2,k_3)$ in which some
$|k_i|\gg 2 \Delta k_f$ are replaced by $U$, the Hubbard constant (this
happens when the initial half band width $\Lambda_0\gg\Delta k_f$, i.e. for
small values of $t_\perp$; it mostly arises from the logarithmic contributions).

\section{Choice of the RG scheme}
\label{choixschema}

It is not the place here to derive the RG equations for the OPI scheme%
\cite{Halboth}, nor for the Wick-ordered one\cite{Nickel}. What matters here is
that one can express the RG flow in terms of couplings $\cal G$, Fermi gap 
$\Delta k_f$, Fermi velocity $\vf$ and renormalization factor $Z$. As far
as we will not  distinguish $v_{f0}$ and $v_{f\pi}$, we need not discuss the
renormalization of $\vf$ and $Z$, which only induces a global scaling of
the other couplings, subsequently we will forget these parameters.

To get the RG equations, one expands diagrammatically all couplings, as in
the Cauchy expansion in $U/\vf$. For a given energy scale $\Lambda$,
one of the inner energies is integrated in the range
$[\Lambda-\delta\Lambda,\Lambda+\delta\Lambda]$, where $\delta\Lambda$ is
infinitesimal; in the OPI scheme, all other inner energies are integrated over
$[\Lambda,\Lambda_0]$; in the Wick-ordered scheme, they are integrated over
$[0,\Lambda]$\cite{Bourbonnais}. OPI and Wick-ordered schemes not only differ
according to these rules, they also give different ${\cal C}$ and ${\cal P}$
factors, as explained before.

From a theoretical point of view, both schemes should converge to the same fixed
point, however, the RG flows are divergent and therefore never reach the
fixed point: the integration of energy is incomplete; therefore, it is
crucial to choose whether one will integrate over UV energies first (i.e.
$|E|>\Lambda$, as in the OPI scheme) or over IR energies first (i.e.
$|E|<\Lambda$, as in the Wick-ordered one)\cite{integration}.

This choice is expected to be more influential when high energy
processes are taken into account. Within the Wick-ordered scheme, such
processes participate in the RG flow at the very beginning, when
$\Lambda\equiv\Lambda_0$, but they are skipped when $\ell$ is increased.
In the OPI scheme, they are always taken into account.

In the ladder system, there is one such process, corresponding to the backward
interband scattering $g_b$. It is indeed a high energy process, only permitted
for $|E|>2\Delta k_f$. Within the Wick-ordered scheme, this contribution is
suppressed for $\ell>\ln(2\vf\Delta k_f/\Lambda_0)$. After these considerations,
one could expect that the RG calculations performed with a fixed Fermi surface
would bring different results, depending on which scheme is chosen.
However, the weight of the $g_b$ contribution is proportional to ${\cal C}$,
${\cal C}^{\rm sp}$, ${\cal P}$ or ${\cal P}^{\rm sp}$. In the OPI scheme, all
these terms vanish as $1/(1+{\Delta k_f\over\Lambda})$, when $\ell$ is
increased, so $g_b$ mostly contributes to the RG flow at the beginning, as in
the Wick-ordered scheme. We have indeed performed both calculations and found a
difference which is meaningless and negligible\cite{unpublished}.

However, if the Fermi surface is correctly renormalized during the RG flow,
in the case when $\Delta k_f\to0$ as far as $\Lambda\to0$, the weight of the
$g_b$ contribution keeps finite during the RG flow, in the OPI scheme, whereas
it is still suppressed for large values of $\ell$ in the Wick-ordered one; so,
the results of RG calculations should prove significantly different, using one
or the other scheme.

In our opinion, the choice of the OPI scheme is more convenient, because, in the
Wick-ordered scheme, the weight of high energy processes is underestimated.
This is, in particular, the conclusion of C. Nickel, who has performed a
careful comparison of different RG schemes (see subsection \textbf{3.4} of
Ref.~\cite{Nickel}).

There is another indication that it is more correct to use the OPI scheme:
in his pioneer work with H. Schulz, D. Zanchi\cite{Zanchi} has
proved that some terms in the 3-loop expansion induce an integration
of energies $|E|>\Lambda$. Therefore, the OPI seems the only self-coherent
scheme, when one tries to include further terms in the perturbative expansion.

\section{Results}
\label{resultats}

\begin{figure}
\epsfysize=10cm
\rotatebox{-90}{\scalebox{0.25}{\includegraphics{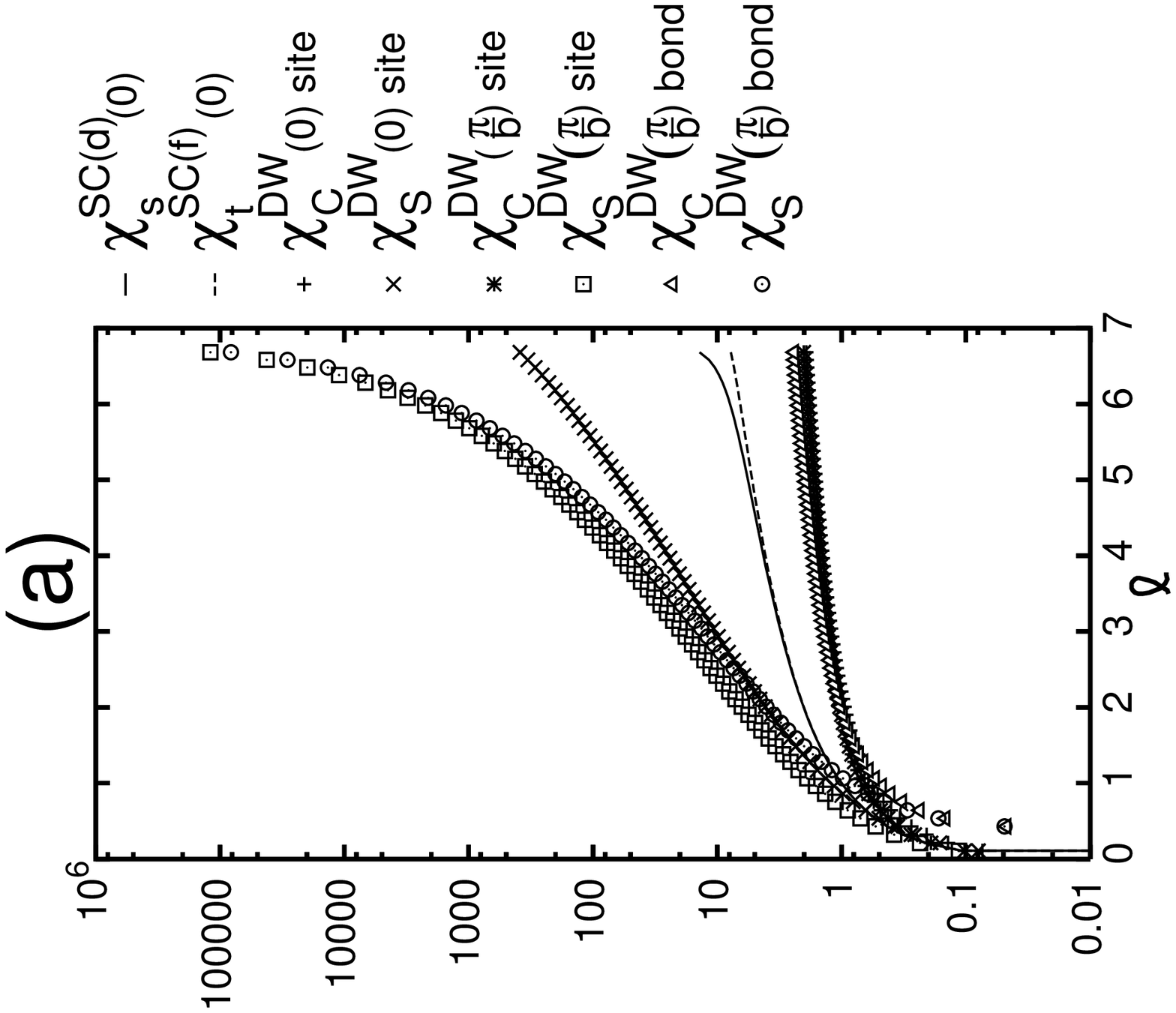}}}
\rotatebox{-90}{\scalebox{0.25}{\includegraphics{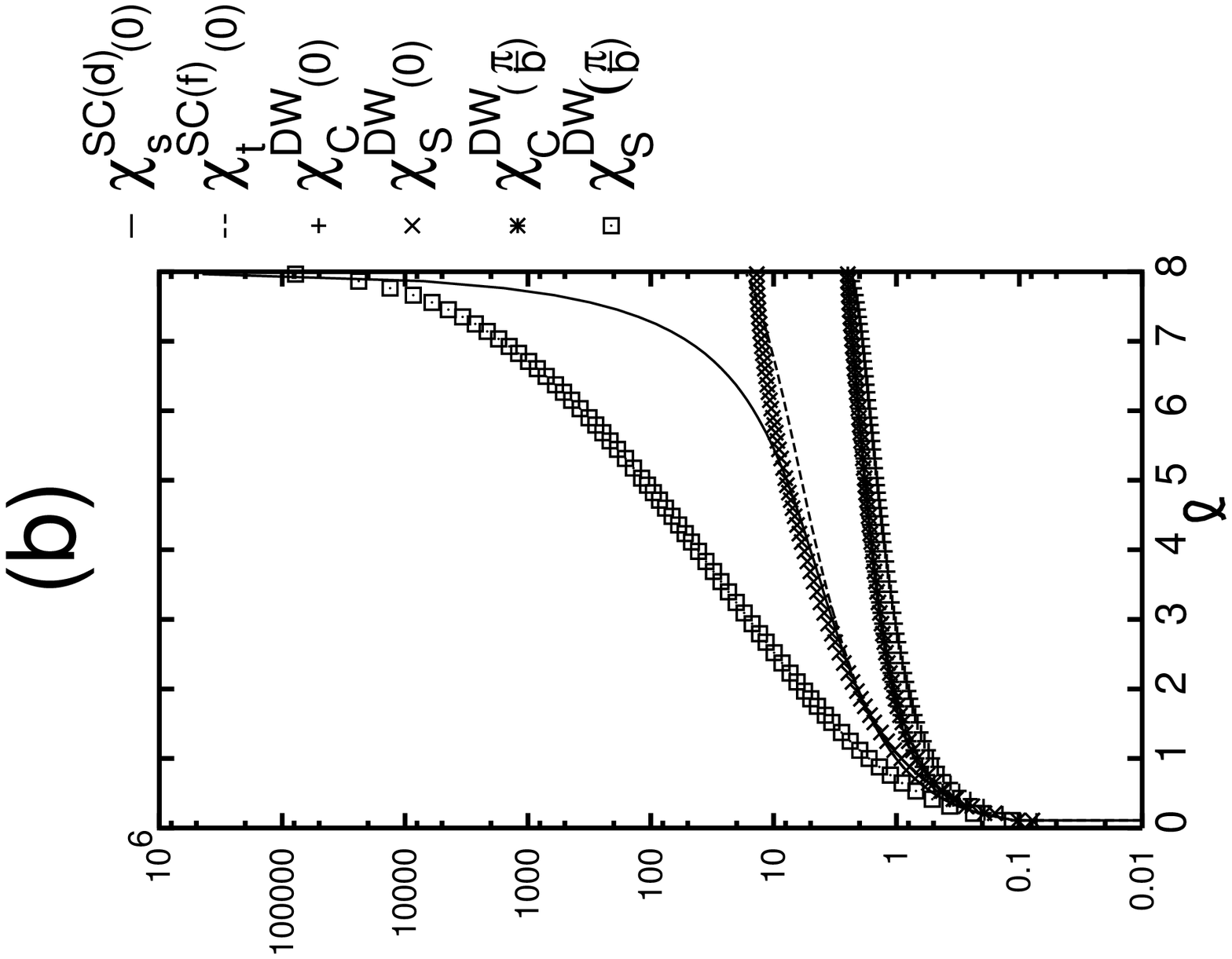}}}
\caption{Flow of the susceptibilities for $2t_\perp/\Lambda_0=1.4$ (a)
or $2t_\perp/\Lambda_0=1.5$ (b), and $\tilde U=0.5$. 
$\chi\lower1.6pt\hbox{${}^{{\rm SC}(d)}_s(0)$}$ is the intraband singular SC
susceptibility of $d$ symmetry, $\chi\lower1.6pt\hbox{${}^{{\rm SC}(f)}_t$}(0)$
is the intraband triplet SC susceptibility of $f$ symmetry, 
$\chi^{\rm DW}_{C}(0)$ is the intraband CDW susceptibility, 
$\chi^{\rm DW}_{S}(0)$ is the intraband SDW susceptibility, 
$\chi^{\rm DW}_{C}({\pi\over b})$ is the interband CDW susceptibility and
$\chi^{\rm DW}_{S}({\pi\over b})$ is the interband SDW susceptibility 
(the difference between site and bond susceptibilities as well as the
symmetry classification are explained in part III of Ref.~\cite{Abramovici}).}
\label{SDWsupra}
\end{figure}

The results of these calculations confirm those of Ref.~\cite{Abramovici}, done
with a fixed Fermi surface (i.e. $\Delta k_f$ was kept constant). We find two
distinct regions; in the SDW region, no superconducting susceptibility is
diverging, while SDW ones are (see Fig.~\ref{SDWsupra}~(a)); in the SC region,
both are diverging, but the superconducting susceptibility always dominates (see
Fig.~\ref{SDWsupra}~(b)). However, the intermediate region, described in
Ref.~\cite{Abramovici}, in which superconducting susceptibilities, although
diverging, are not dominating, vanishes completely.

\begin{figure}
\epsfysize=10cm
\rotatebox{-90}{\scalebox{0.30}{\includegraphics{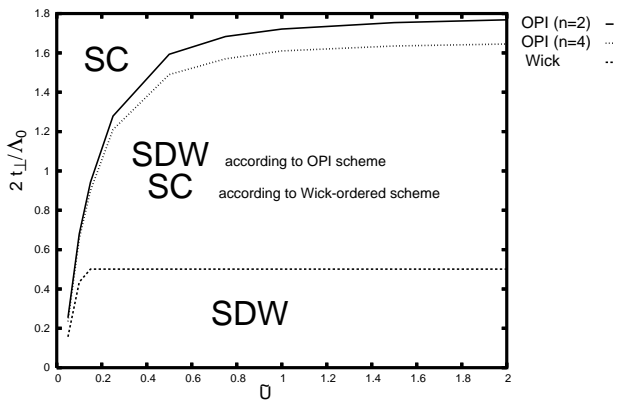}}}
\caption{Phase diagram, versus parameters $t_\perp$ and
$\tilde U={U\over\pi \vf}$; the central area belongs to the SDW phase, according
to the OPI scheme, and to the SC one, according to the Wick-ordered scheme.
The parameter $n$ indicates range $[-2n\Delta kf,2n\Delta k_f]$ in which
$K_\Vert$ dependency is exactly taken into account. We did not distinguish
the curves for $n=2$ and $n=4$ in the Wick-ordered scheme, because they hardly
separate {\it de visu}.}
\label{phasdiag}
\end{figure}

In Fig.~\ref{phasdiag}, the two phase diagrams are presented, according to the
choice of the RG scheme. Let us repeat that, when $\Delta k_f$ is not
renormalized, both schemes give almost the same phase diagram\cite{unpublished}.
Here, on the contrary, one observes that the SDW region quantitatively depends
on the RG scheme. Indeed, for all values of $U$, except very small ones, in the
Wick-ordered calculation, a constant critical value $t_{\perp c}$ can be
defined, which separates the SDW and the SC regions. In the OPI calculations,
the evolution of this critical value $t_{\perp c}$ is smoother, with a linear
part of slope $\sim8.6$ at small $U$. On the whole, the critical line
$t_{\perp c}(U)$ which separates both regions differs quantitatively, except for
small values of $U$\cite{zeroimprecis}. With the Wick-ordered RG scheme, the SDW
area is reduced by a factor 3, compared to the result of the RG with the OPI
scheme.

The difference of results coming from the choice of the RG scheme has already
been suggested by several authors. It was, in particular mentioned in
Ref.~\cite{Rohe}. 

The behaviour of the Fermi surface, during the RG flow, brings no surprise. Let
us first present the results in the SC phase, then in the SDW one.

In the SC phase, $\Delta k_f$ increases slowly, while not diverging. The flow
diverges at some $\Lambda_c$, and $\Delta k_{fc}$ is the final value of
$\Delta k_f$. The numerical values of $\Delta k_{fc}/\Lambda_0$ are not
realistic, however the general trend is very satisfactory and indicates that the
binding/antibinding separation is necessary to the existence of
superconductivity.  We believe that, if one would include a greater number of
chains in the model, one would obtain more realistic values for $\Delta k_{fc}$.

Let us emphasize the importance of a non-zero value of $\Delta k_f$. As
discussed by Clarke, Strong and Anderson\cite{Clarke}, the properties of
Luttinger liquid, which have been established for a single
one-dimensional chain, can extend in the case of a quasi-one-dimensional
system (spin/charge separation, power-law behaviour of correlation functions);
that is, even though the band structure extends in the $\perp$ dimension (as
$\Delta k_f\ne0$), the system will not converge to the two-dimensional Fermi
liquid. These authors claim that $\perp$ superconductivity originates from
this mechanism, which also relates to confinement in the $\Vert$ dimension.
From this point of view, unconventional
superconductivity and Luttinger liquid concept (in particular spin/charge
separation) are interplaying; this gives an explanation for the possibility
of coexistence of SDW instabilities and superconductivity.

\begin{figure}
\epsfysize=10cm
\rotatebox{-90}{\scalebox{0.16}{\includegraphics{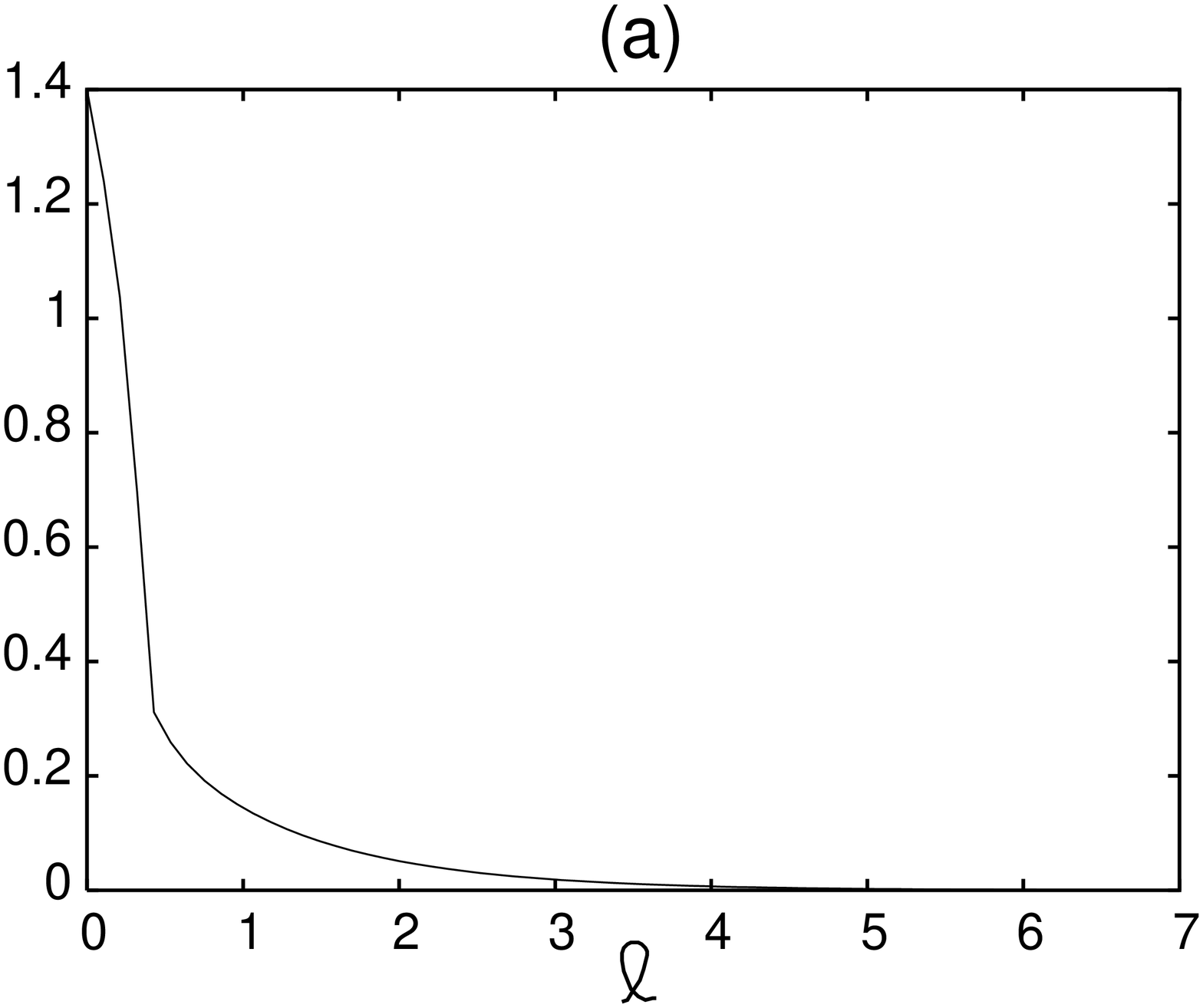}}}
\rotatebox{-90}{\scalebox{0.16}{\includegraphics{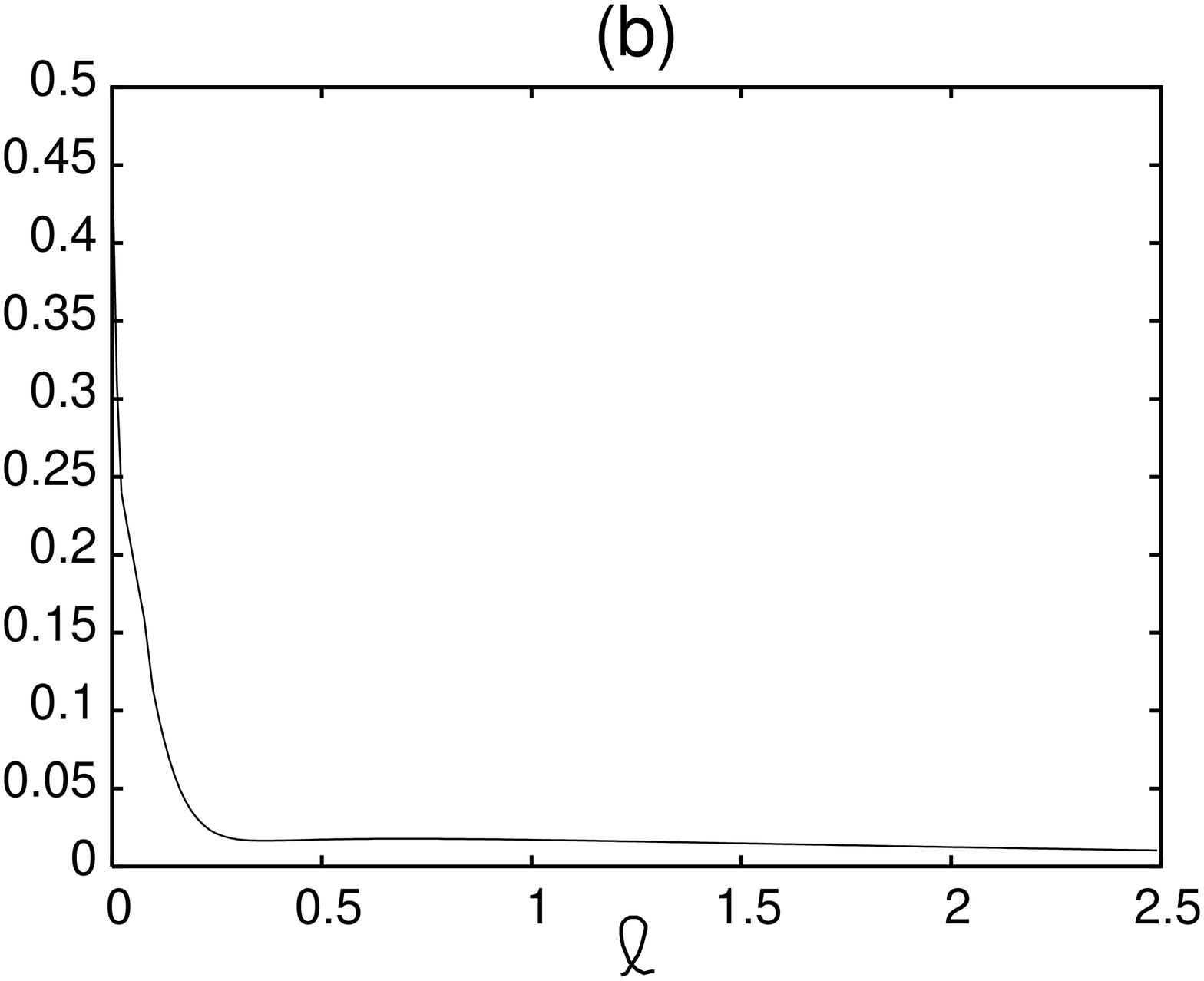}}}
\rotatebox{-90}{\scalebox{0.16}{\includegraphics{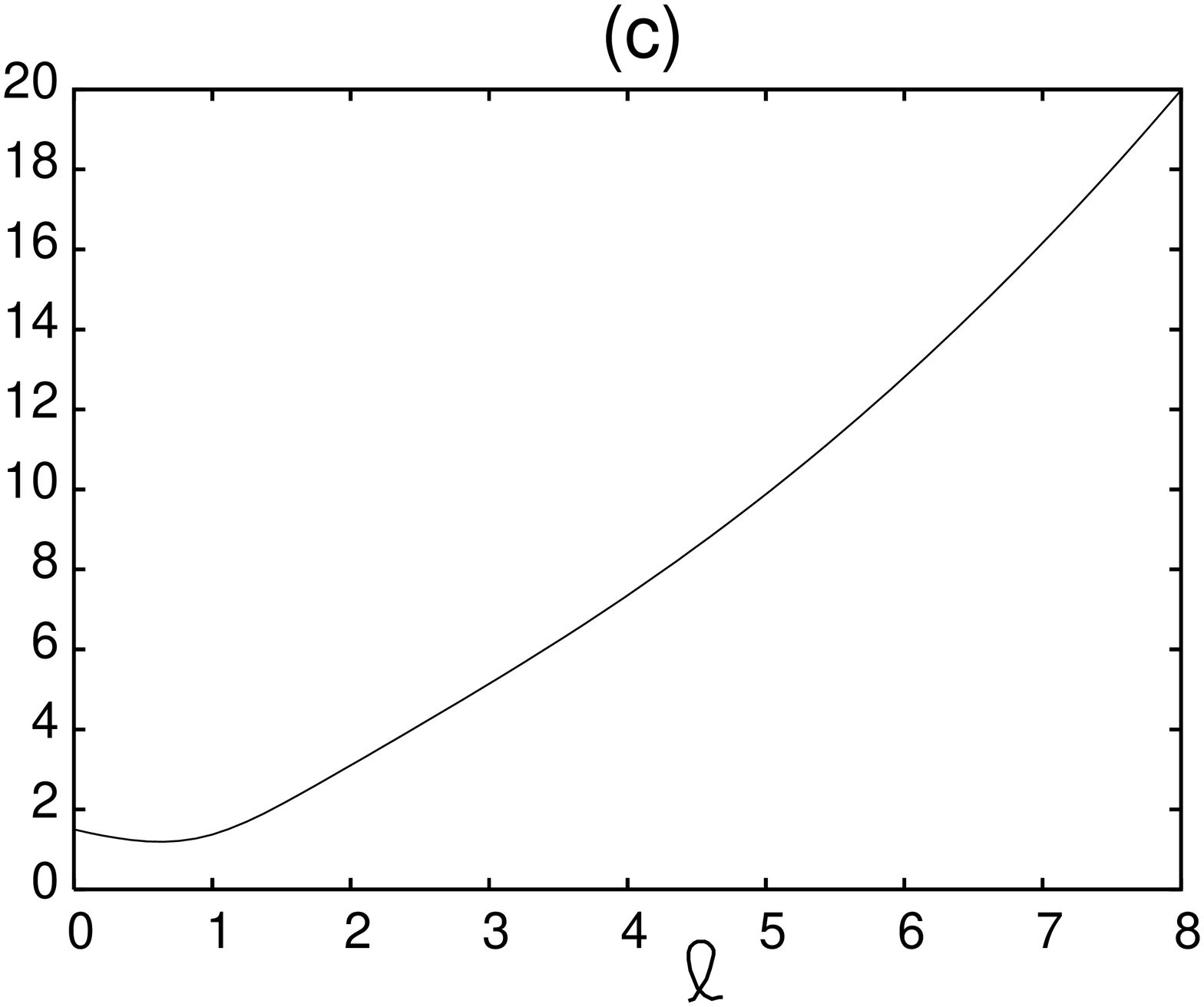}}}
\rotatebox{-90}{\scalebox{0.16}{\includegraphics{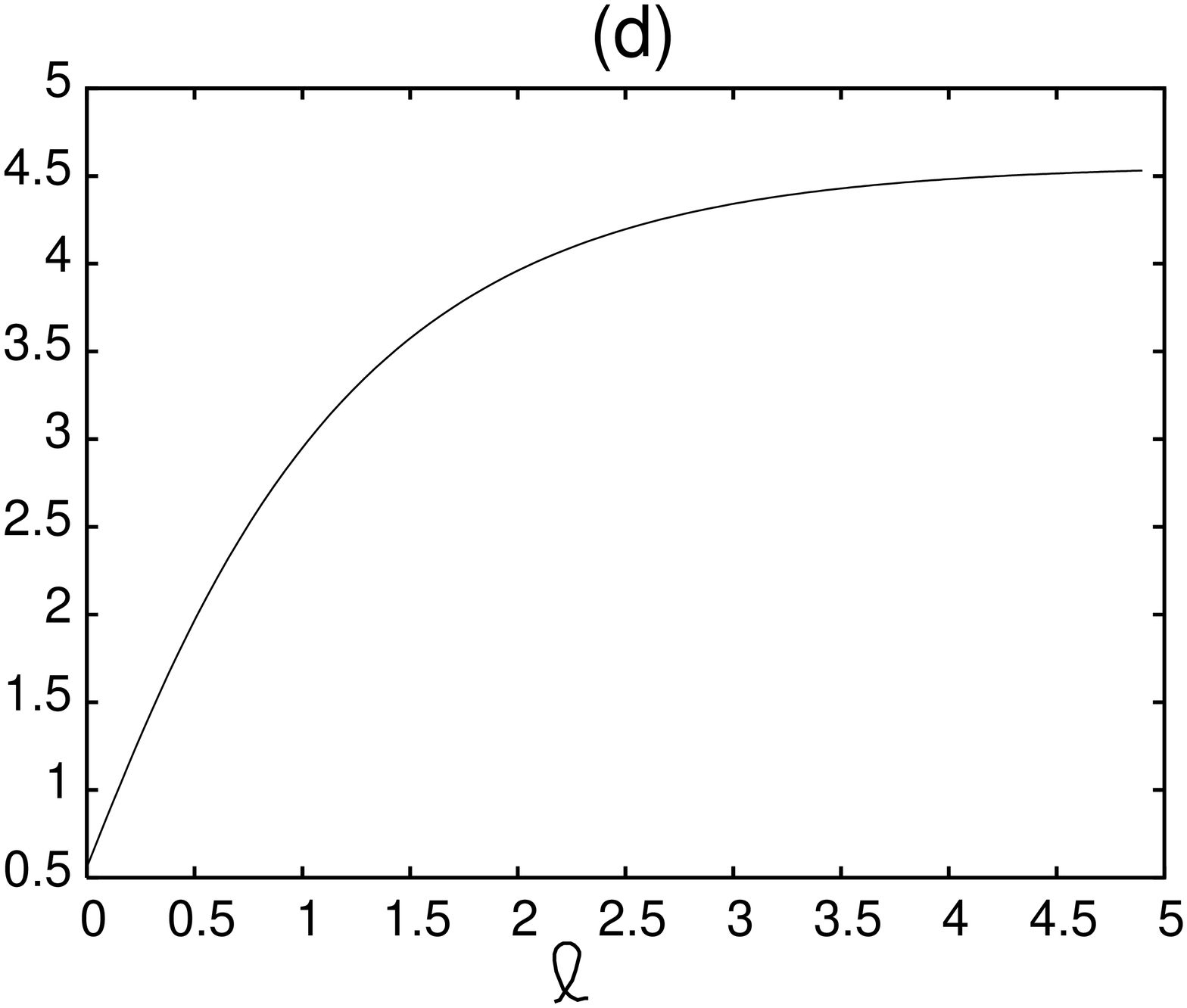}}}
\caption{Flow of the $\Delta k_f$: in the SDW phase,
(a)~using OPI RG calculations, or (b)~using Wick-ordered RG calculations;
in the SC phase, (c)~using OPI RG calculations, or
(d)~using Wick-ordered RG calculations.}
\label{fermi}
\end{figure}

In this SC phase, we also observe a quantitative difference between the results
obtained using a OPI or a Wick-ordered scheme. In the first case, the value of
$\Delta k_f$ lies in the interval $[25,30]$, while in the second, it lies in
$[4,5]$ (see Fig.~\ref{fermi} (c) and (d)).

\begin{figure}
\epsfysize=10cm
\rotatebox{-90}{\scalebox{0.27}{\includegraphics{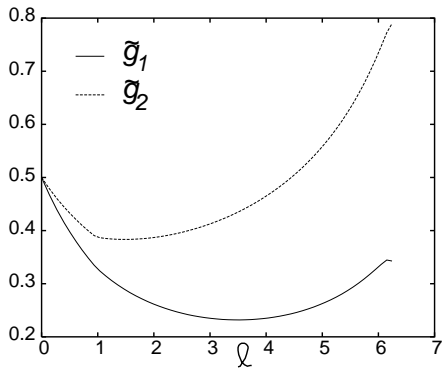}}}
\caption{Flow of couplings $g_{01}$ and $g_{02}$ during the RG flow, for
$\tilde U=0.5$ and $2t_\perp/\Lambda_0=1.485$.}
\label{gg}
\end{figure}

In the SDW phase, $\Delta k_f\to0$ as $\Lambda\to\Lambda_c$ (see
Fig.~\ref{fermi} (a) and (b)). This proves that this phase relates to the
Luttinger solution. Contrary to the SC phase, the band structure remains
purely one-dimensional. This system, however, is different from Luttinger's
original one, because, in real space, there are still two chains, with
non-zero hopping in-between. Let us examine in detail the  behaviour of the
scattering susceptibilities. Couplings $g_{01}$ and $g_{02}$ are not diverging
(see Fig.~\ref{gg}), contrary to what is observed in the SC phase. The curves of
all couplings are very close to those obtained when $\Delta k_f$ is kept
constant. These results are very different from that of Fabrizio, who finds
the behaviour of a single chain (see Ref.~\cite{Fabrizio} and compare with
Fig.~\ref{gg} here).

One must observe that, as $\Delta k_f\to0$, the effective range of $K_\Vert$
shrinks, so that, when one approaches the critical scale $\Lambda_c$, the
$K_\Vert$ dependence is extremely badly taken into account. On could even expect
to recover usual RG calculations, for which no SDW phase is found; however,
the observation of this phase has proved surprisingly robust; the explanation is
probably that, just before $\Lambda_c$, the divergence of $\chi^{\rm DW}_S$
dominates already in such a way that it prevents any divergence of
$\chi\lower1.6pt\hbox{${}^{{\rm SC}(d)}_s$}$. Nevertheless, this discussion
sheds also light about serious numerical convergence problems that arise
in this region, and have required technical answers.

Let us compare these calculations with experimental data. No direct
determination of $U$ or $t_\perp$ are available, one can only get indirect 
determinations by matching experimental and theoretical curves, as it is
done in Ref.~\cite{Tsuchiizu}, for $Sr_{14-x} Ca_x Cu_{24} O_{41}$ compounds
(with $x=12$). These authors  compare several experimental and theoretical
spin susceptibility curves (including uniform spin susceptibilities) and obtain
a best fit for $U/t_\Vert\sim4$ and $|t_\perp|\sim t_\Vert$ (those values
correspond here to $\tilde U\sim 1$ and $2t_\perp\sim\Lambda_0$), which
corresponds here to predictions done with the Wick-ordered scheme; other
determinations are available, which correspond to the OPI scheme. Actually, the
determinations are accurate within an order of magnitude, thus it is not 
possible, from experimental data, to decide which scheme gives the right
predictions. However, many theoretical predictions, and a few experimental fits,
are located in this region of parameters, close to point
$(\tilde U\sim1,2t_\perp/\Lambda_0\sim1)$ in the phase diagram (for instance
by a factor $~2$). In particular the boundary, between SC and SDW phases, is
located in the same region of parameters. Therefore, even if we can't
discriminate between Wick-ordered and OPI schemes, the phase diagrams we have
calculated are qualitatively in good agreement with experimental observations.

In conclusion, we would like to emphasize that these calculations confirm the
determination of a pure SDW phase using a very simple ladder model, which was
far from being obvious until now. The SC phase also indicates a possible
coexistence of magnetism and superconductivity, as it is indeed observed,
both theoretically and experimentally. 

We have also established the importance of the choice of the RG scheme. Even if
this alternative only raises quantitative differences, they are not negligible,
so this has to be carefully taken into account. We hope that in further and more
precise models, a clear discrimination between the two schemes will be possible,
and that it will confirm our conjecture that the OPI scheme is more accurate.

The authors would like to thank deeply Christoph Nickel for his
collaboration in the theoretical preparation of this paper; without his help,
this work would probably never have been completed successfully.


\begin{thebibliography}{fe}
\bibitem{integration} In one-loop diagrams, there is only one integration,
however, the domain of external momenta are different in the OPI scheme and in the
Wick-ordered one.
\bibitem{unpublished} All calculations in Ref.~\cite{Abramovici} are done using
the OPI scheme. They have been also performed using the Wick-ordered scheme, but
there are no noticeable differences.
\bibitem{zeroimprecis} It is a fact that all curves are exactly converging to
the point ($2t_\perp/\Lambda_{\rm o}=0.02,U=0$), however, one has to take these
extrapolations very carefully, because the theoretical precision of the
calculations behaves like $\Lambda_c/\Lambda_{\rm o}$ and one observes that
$\log(\Lambda_c/\lambda_{\rm o})>100$ for $2t_\perp/\Lambda_{\rm o}\le0.01$.
\end{thebibliography}
\end{document}